\UseRawInputEncoding
\documentclass[aps,prc,twocolumn,showkeys,showpacs]{revtex4}
\usepackage{graphicx}
\usepackage{dcolumn}
\usepackage{color}
\usepackage{bm}

\begin{document}
\newcommand{\bel}[1]{\begin{equation}\label{#1}}
\newcommand{\belar}[1]{\begin{eqnarray}\label{#1}}
\def\bra{\langle}
\def\ket{\rangle}
\def\eps{\epsilon}
\def\epsk{\epsilon_k}
\def\mev{\;{\rm MeV}}
\def\tc{\textcolor{red}}
\def\tb{\textcolor{blue}}
\def\pr{\prime}
\def\dpr{\prime\prime}

\title{Shell effects and multi-chance fission in the sub-lead region}

\author{ F. A. Ivanyuk}
\email{ivanyuk@kinr.kiev.ua}
\affiliation{Institute for Nuclear Research, 03028 Kiev, Ukraine}

\author{C. Schmitt}
\email{christelle.schmitt@iphc.cnrs.fr}
\affiliation{IPHC, University of Strasbourg, 67200 Strasbourg, France
\footnote{\flushbottom Present address: Henryk Niewodniczanski Institute of Nuclear Physics PAN, 31-342 Krakow, Poland}}

\author{C. Ishizuka}
\email{chikako@zc.iir.titech.ac.jp}
\affiliation{Institute of Science Tokyo, Tokyo, 152-8550 Japan}
\author{S. Chiba}
\email{absmithjp@gmail.com}
\affiliation{NAT Corporation, 38 Shinko-cho, Hitachinaka City, Ibaraki Prefecture, 312-0005, Japan}

\date{today}

\begin{abstract}
Within the recently developed  five-dimensional Langevin approach for the description of fission of heavy nuclei, we have calculated the fission fragments mass and kinetic energy distributions for the fission of  $^{180}$Hg and $^{190}$Hg formed  in the reactions $^{36}$Ar +$^{144}$Sm $\Rightarrow$ $^{180}$Hg and
$^{36}$Ar +$^{154}$Sm $\Rightarrow$ $^{190}$Hg at few excitation energies and found very good agreement between the calculated and experimental  results. Special attention was paid to the accurate description of the dependence of shell effects on the excitation energy. It was shown that the effect of multi-chance fission on the mass distribution is noticeable only at small excitation energies. The kinetic energy distributions are more sensitive to pre-scission neutron emission.
\end{abstract}

\pacs{24.10.-i, 25.85.-w, 25.60.Pj, 25.85.Ca}
\keywords{nuclear fission, Langevin approach, fission fragments mass distributions, shell effects}

\maketitle

\section{Introduction}
\label{intro}
Since its discovery in the late 1930s \cite{HS,meitner,wheeler}, nuclear fission has been extensively studied both theoretically and experimentally. Nuclear fission is still a mysterious phenomenon, characterized as a complex large-amplitude collective motion of a finite number of nucleons influenced by both liquid-drop behavior and shell structure.

Among these studies, the mass number distribution of fission fragments and the total kinetic energy (TKE) exhibit both clear systematic trends and anomalies that deviate from these trends. These include the systematic double-humped mass distribution in the actinide region driven by shell effects of $^{132}$Sn and deformed shell in the region around A=142, the abrupt change to a
sharply single-humped distribution in the Fm region, and the change from a single-humped distribution dominated by liquid-drop
behavior in the pre-actinide region to an unexpected double-humped distribution in $^{180}$Hg \cite{Andreev2010}. Such systematic trends and anomalies are intricately intertwined. In the superheavy element region, there may also exist a super-asymmetric mode driven by the shell effects of $^{208}$Pb \cite{Ishizuka2020,pasha,BSM,poenaru2011,poenaru2018,warda2018,cluster}.

We have previously used a fluctuation-dissipation mo\-del based on the four-dimensional Langevin equations \cite{our17,scirep} to study these aspects of nuclear fission, successfully elucidating the mechanisms of systematic trends and anomalies from the actinide to the superheavy element region. Recently, we have extended this model to five dimensions (5D) \cite{5dimen} and have been able to describe the emergence of a triple-humped structure in the thorium region.

In this study, we apply the five-dimensional Langevin method to quantitatively reproduce
the emergence of a double-humped mass distribution in $^{180}$Hg observed in \cite{Nishio2015}. The study aims to
explore the mechanism behind this phenomenon. To this end, we consider fission of $^{180}$Hg and $^{190}$Hg
induced in the reactions $^{36}$Ar +$^{144}$Sm $\Rightarrow$ $^{180}$Hg and $^{36}$Ar +$^{154}$Sm $\Rightarrow$ $^{190}$Hg \cite{Nishio2015}, and compare the
calculated fragment mass and total kinetic energy (TKE) distributions with the experimental data at
several bombarding energies.

Since the surprising result of Andreyev et al \cite{Andreev2010}, theoretical work on fission of nuclei in the sub-lead region
\cite{Moller2012,Ichikawa2012,Warda2012,Sida2012,McDon2014,Scamps2019,Ichikawa2019,Pasca2020,Pomorski2020,Huo2024}, has been restricted to either static analyses of potential energy surfaces from various models
(macroscopic-microscopic and self-consistent) leading to 'hand waving' conjectures about the shape
of the mass distribution, scission point models, the WKB+Born-Oppenheimer approximation, the
Brownian shape motion model, or self-consistent TDHF restricted to the very last stage of the saddle-to-scission descent. To the best of our knowledge, no fully dynamical model of fission 
starting from the equilibrium shape up to scission has been applied to
the sub-lead region so far. We undertook the challenge using our five-dimensional Langevin code with
the aim to understand fission in the region from a unique perspective. It must be noticed that
we do not assume any
distribution as the initial condition; our calculation starts from delta-function in the 5 variables.
This means that the distributions in the mass and TKE themselves arise as results
 of shape evolution described by Langevin equations.

Besides, in the present work we tried to clarify the role of shell effects the influence of which due to multi-chance fission is seen at rather high  excitation energies, 30-50 MeV, observed e.g. in \cite{Nishio2015,ryzhov,simutkin}.

The importance of shell effects in  nuclear fission is well established in numerous experiments and described by many theoretical works. The delicate question here is the dependence of shell effects on the excitation energy. The nuclear deformation energy is most often described within the macroscopic-microscopic models as the sum of the macroscopic (liquid-drop) energy and the temperature-dependent shell correction. The calculation of the shell correction at finite temperature is quite involved. So, as a rule, the shell correction $\delta E$ is calculated at zero temperature and for the dependence of $\delta E$ on the excitation energy some approximation is used.

According to a very popular prescription by A. Ignat\-yuk \cite{ignat} the shell correction decreases exponentially with growing excitation energy $E^*$, $\delta E(E^*)=\delta E(0)\times \exp{(-E^*/\gamma)}$ with the  damping factor $\gamma \approx$ 20 MeV. At the same time there is information from the experiments \cite{Nishio2015,ryzhov,simutkin} that the shell effects in fission are seen at rather high excitation energies, $E^*=$ 30-50 MeV. That is why, in the present work we have put special attention on the dependence of shell corrections on the excitation energy.

It turned out that the calculated first chance fission results for the fission fragment mass distributions are very close to the experimental data.
Then we took into account the multi-chance fission. The agreement with the experiment became even better, especially for the TKE distributions.

Describing fission of actinides (like many models are able to do nowadays) is not sufficient
to reveal the multiple facets of the process. A consistent description in other regions of the
nuclear chart, and in particular in the sub-lead corner which has shown to exhibit rather
different features as compared to actinides, is mandatory.

In Section \ref{langevin} we present the main relations of the Langevin approach to the description of nuclear fission process.  In Section \ref{poten} we explain how the collective potential energy is calculated. Section \ref{transp} contains the description of the used mass and friction tensors. The results of numerical 5D calculation are given in Section \ref{results}.
The account of multi-chance fission with neutron emission is presented in Section \ref{multi}.
 Section \ref{summa} contains a short summary.

\section{The Langevin approach}
\label{langevin}
In the Langevin approach one solves the set of differential equations for the time evolution of collective variables $q_{\mu}$
describing the shape of the nuclear surface. For the shape
parametrization we use in our works that of the two-center shell model (TCSM) \cite{tcsm}. In this model the shape of the axially symmetric nucleus is characterized by 5 deformation parameters $q_{\mu} =z_0/R_0, \delta_1, \delta_2, \alpha $ and $\epsilon$.
Here, $z_0/R_0$ refers to the distance
between the centers of left and right oscillator potentials, with $R_{0}$ being the radius of the spherical nucleus. The parameters $\delta_1$ and $\delta_2$ describe the
deformation of the right and left parts of the nucleus. The fourth parameter
$\alpha $ is the mass asymmetry and the fifth parameter $\epsilon$ of TCSM shape
parametrization regulates the neck radius. All details about the shape variables used in the present work 
can be found in Ref.\cite{5dimen}.

\begin{figure}[ht]
\centering
\includegraphics[width=0.85\columnwidth]{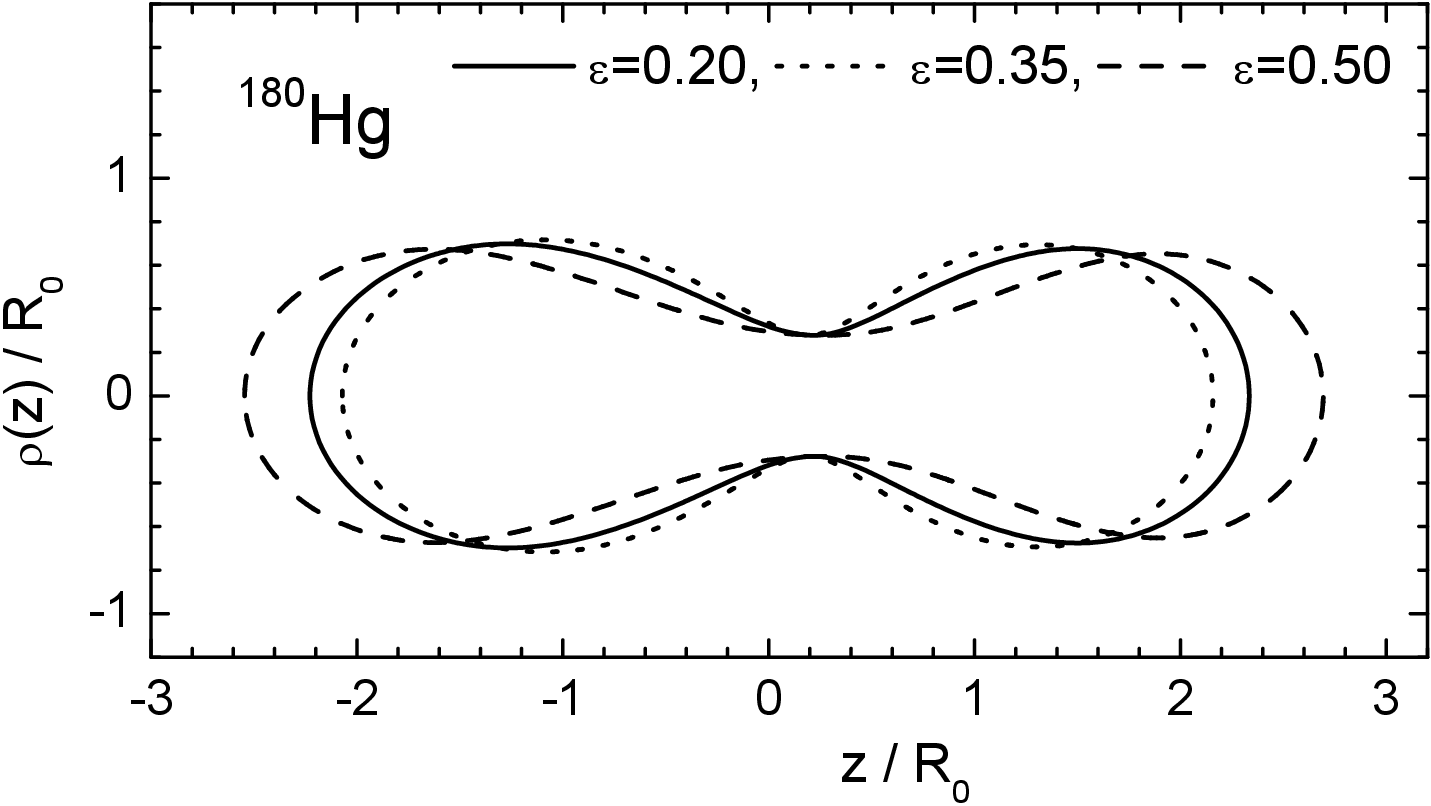}
\caption{The TCSM shapes for $\delta_1=\delta_2=\alpha$=0.2 at the scission point for few values of parameter $\eps$}
\label{shapes180}
\end{figure}
In several of our previous works, the parameter $\epsilon$ was kept constant, and equal to 0.35. 
In a recent publication \cite{5dimen} we considered $\epsilon$ as a dynamical variable, similar to the other four deformation parameters. Calculations within the 5D 
dynamical Langevin approach provides a better description of the experimental data, as compared with 4D calculations. In particular, the transition from mass-symmetric to mass-asymmetric fission via the triple-humped distribution in fission along the thorium isotopic chain can be reproduced, see Ref. \cite{5dimen}.

Formally, the parameter $\epsilon$ controls the neck radius. However,
since the fragment properties are calculated at the scission point defined with a fixed neck radius, the variation of $\epsilon$ gives access to very elongated or compact shapes, see Fig. \ref{shapes180}. Such shapes contribute to the super-long or super-short fission modes, respectively \cite{scirep}.

The first-order differential equations (Langevin equations) for the time
dependence of the collective variables $q_{\mu }$ and the conjugate momenta
$p_{\mu }$ are:

\begin{eqnarray}\label{lange}
\frac{dq_\mu}{dt}&=&\left(m^{-1} \right)_{\mu \nu} p_\nu , \\
\frac{dp_\mu}{dt}&=&-\frac{\partial F(q,T)}{\partial q_\mu} - \frac{1}{2}\frac{\partial m^{-1} _{\nu \sigma} }{\partial q_\mu} p_\nu p_\sigma\nonumber\\
 &-&\gamma_{\mu \nu} m^{-1}_{\nu \sigma} p_\sigma+g_{\mu \nu} R_\nu (t),
\end{eqnarray}
where the sums over the repeated indices are assumed. In Eq.(2) $F(q, T)$ is the
temperature dependent free energy of the system, $\gamma _{\mu \nu }$
and $(m^{-1})_{\mu \nu }$ are the friction and inverse of mass tensors, and $g_{\mu \nu}R_{\nu}$(t) is the random force.

The free energy $F(q, T)$ is calculated as the sum of the macroscopic (folded Yukawa) energy
and the temperature dependent shell correction $\delta F(q, T)$.
The shell corrections at zero temperature are calculated from the single-particle energies in the deformed
Woods-Saxon potential
\cite{pash1,pash2} fitted to the mentioned above TCSM shapes.



The configuration space for solving Eqs. (1, 2) should be large enough to include all shapes that can be populated during the fission process. In the present work we used $0.08\leq z_0/R_0\leq 4.28$, $\Delta z_0/R_0=0.105$, $-0.75\leq\delta_1, \delta_2\leq 0.75$, $\Delta\delta_{1,2}=0.05$, $-0.78\leq \alpha\leq 0.78$, $\Delta\alpha=0.04$, $0.1\leq \eps\leq 0.9$, $\Delta\eps=0.1$.

During the time evolution the trajectory may reach the boundary of the deformation space. If one were to  abandon such trajectory very few trajectories would reach the scission region and the calculations would take a lot of time. The popular method to keep the trajectory inside the deformation space is to change the sign of the collective velocity in the direction of the boundary when the trajectory 'hits' the boundary or to put the walls at the boundaries. Both methods are not perfect. In both methods the trajectory receives a 'kick' from the boundaries that is not consistent with Eq. (\ref{lange}) and energy conservation. The trajectories gain additional kinetic energy from the boundary, the temperature becomes negative very soon and such trajectory is stopped.

A more accurate method of treating the collisions with the boundaries was proposed by A. Sierk \cite{sierk}. He proposed to put a 'soft wall' at the boundaries, which is a potential $V(x)$ that is parabolic at small (positive) distance $x$ to the boundary and rise exponentially at large  distances from the boundary:
\begin{equation}\label{softw}
V(x)=V_{wall}[\exp(x)-1.0-x] ,
\end{equation}
where $V_{wall}$ is a constant of the order of 100 MeV, see Fig. \ref{reject}.
\begin{figure}[ht]
\centering
\includegraphics[width=0.8\columnwidth]{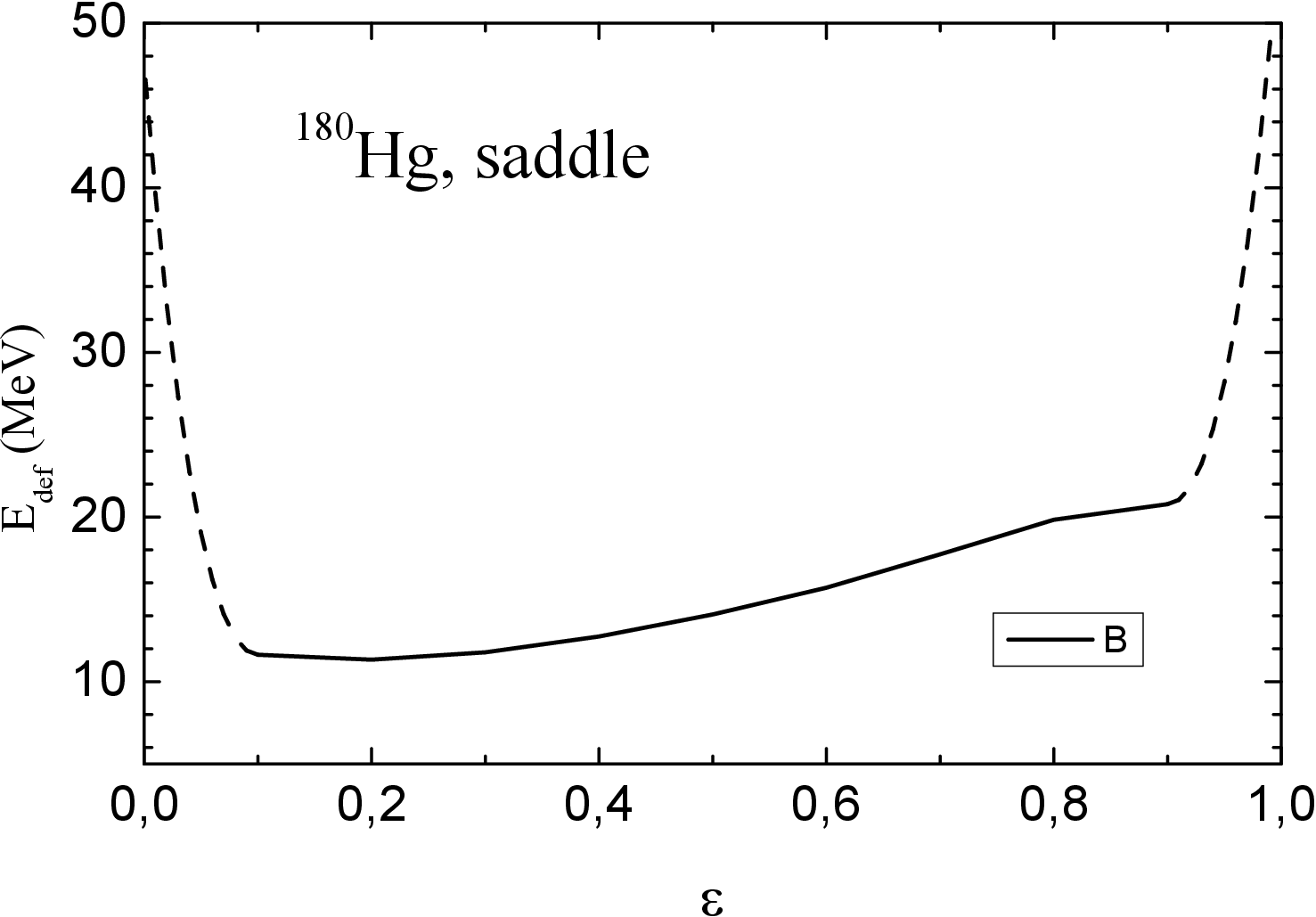}
\caption{The dependence of the potential energy (\ref{free})
of $^{180}$Hg  at $T=0$  on the deformation parameter $\eps$ at the saddle point is shown by the solid line,and the 'soft wall' (\ref{softw}) with $V_{wall}$=100 MeV is shown by the dashed line.}
\label{reject}
\end{figure}
By moving in the potential (\ref{softw}) the energy is conserved. Therefore, the trajectory is rejected by the boundary with the same energy. This method keeps the trajectories within the boundaries for a long time.  Many trajectories have a chance to reach the scission point, thus making the computations faster which is very important for the very time-consuming 5D calculations.

The collective inertia tensor $m_{\mu \nu}$ is calculated within the
Werner-Wheeler approximation \cite{werwhe}
and for the friction tensor $\gamma_{\mu \nu }$ we
used the wall-and-window formula  \cite{wall1,runswiat}.

The random force $g_{\mu \nu}R_{\nu }(t)$ is the product of the normally distributed white noise
$R_{\nu}(t)$, $\langle R_{\mu}(t)R_{\nu}(t^{\pr})\rangle=2 \delta_{\mu\nu}\delta (t-t^{\pr})$, and the temperature-dependent strength factors
$g_{\mu \nu}$. The factors $g_{\mu \nu }$ are related to the temperature
and friction tensor via the modified Einstein relation,
\begin{equation}\label{teff}
g_{\mu \sigma } g_{\sigma \nu } =T^\ast \gamma _{\mu \nu }
\,,\,\,\,\,\,\,\,\,\,\,\,\,\,\,\,\,\,\,\,\,\,\,T^\ast
=\frac{\hbar \varpi}{2}\coth \frac{\hbar \varpi}{2T}\,\,,\,\,\,\,
\end{equation}
where the effective temperature $T^{\ast }$ is defined in the references \cite{pomhof,hofkid}. The parameter $\varpi$
 represents the local frequency of collective motion \cite{hofkid}. The minimum value of
$T^{\ast }$ is determined by $\hbar \varpi/2$. For large excitations $T^{\ast }$ approaches $T$.

In the present work we used for $\varpi$ the deformation-independent value $\hbar\varpi$=2 MeV.

The temperature $T$ in (\ref{teff}) is related to the initial energy $E^{\ast}_{(in)}$ and the local excitation energy $E^{\ast}$ by
\begin{equation}\label{Exxx}
E^\ast =E_{gs} +E^\ast _{(in)} -\frac{1}{2}m^{-1}_{\mu \nu } p_\mu p_\nu
-V_{pot} (q,T=0)=aT^2,
\end{equation}
where $V_{pot}$ is the potential energy and $a$ is the level density parameter.
For the level density parameter we use the approximation of \cite{wada93}
\begin{equation}\label{alevel}
 a = A (1.0+3.114 A^{-1/3}+5.626 A^{-2/3}) / 14.61\mev,
\end{equation}
were the mass number is denoted by $A$.
More details can be found in our earlier publications, see  \cite{scirep,24,four,26}.

Usually, the initial  values of momenta $p_{\mu}$ are set to zero, and calculations are
started from the ground state deformation.
The calculations are continued until the trajectories reach the "scission point". In our older publications the "scission point" was defined as the point in deformation space where the neck radius becomes zero. However, the zero critical neck radius is not well justified. It was shown in the so called optimal shapes model \cite{strut1963,pomiva2} that the nuclear liquid drop loses stability with respect to elongation at an almost constant elongation $z_0\approx$ 2.3 $R_0$ and rather thick neck, $r_{neck}^{crit}$. Thus, in all our calculations below we used the finite value of the critical neck, $r_{neck}^{crit}$ = 2 fm. For each system, five millions of trajectories were run.

\section{The temperature dependence of the potential energy}
\label{poten}

The potential energy $F(q)$ is calculated within the macroscopic-microscopic model,
\begin{equation}\label{free}
F(q)=E_{LDM}(q)+\delta F(q,T)
\end{equation}
\begin{figure}[ht]
\centering
\includegraphics[width=0.95\columnwidth]{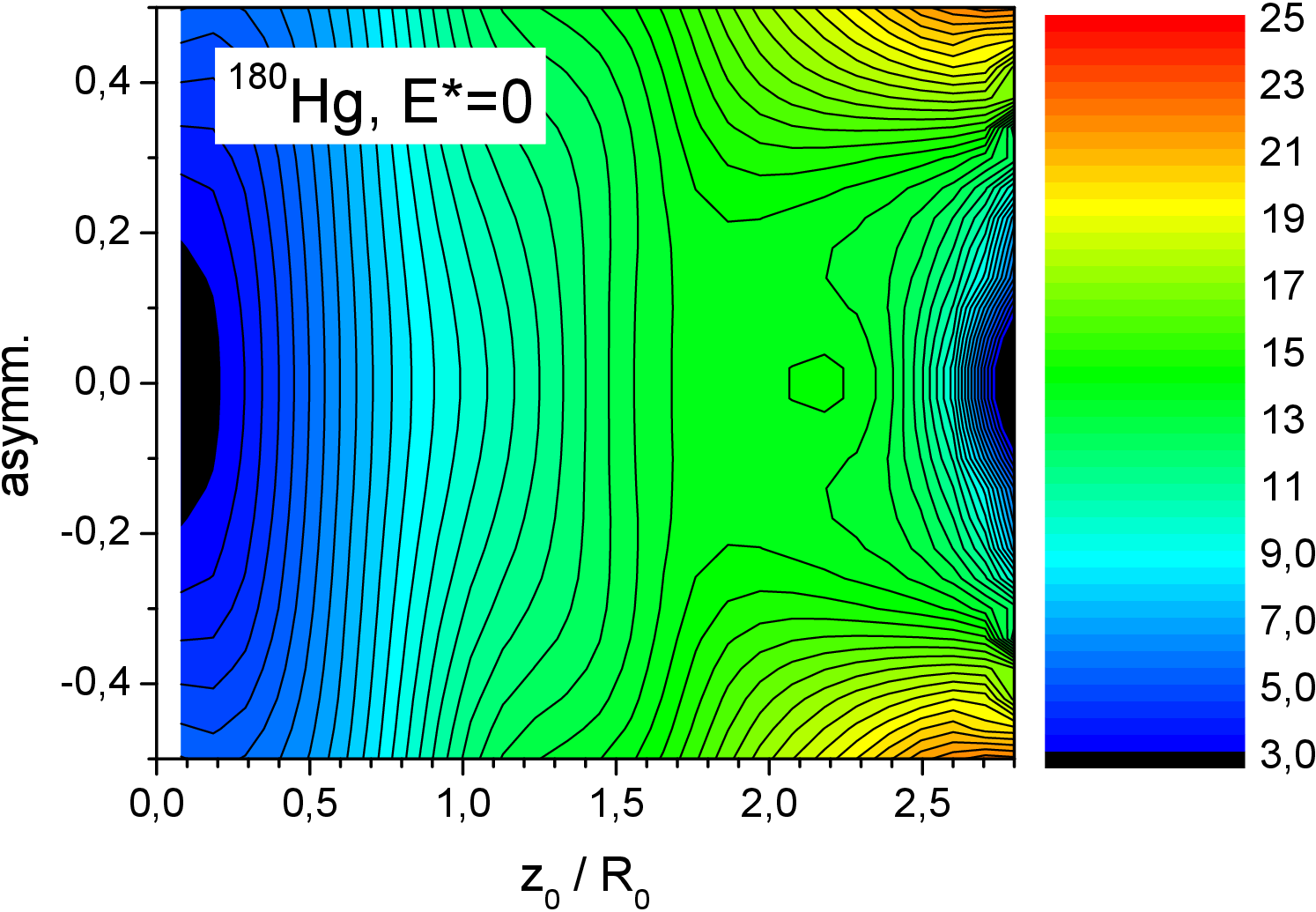}
\caption{(color online) The dependence of the mean value of the potential energy $\bra F \ket$ (\ref{free}) of $^{180}$Hg on the elongation $z_0/R_0$ and the mass asymmetry $\alpha$.}
\label{edef180}
\end{figure}

The macroscopic part of the energy $E_{LDM}(q)$ is calculated within the folded Yukawa model \cite{suek74,iwam76,sato79}.
At zero temperature the shell correction to the free energy $\delta F(q,T=0)$ coincides with the shell correction to the collective potential
energy $\delta E(q)$. The shell correction $\delta E(q)$ is calculated by Strutinsky's prescription \cite{struti,brdapa} from the energies of single-particle states in the deformed Woods-Saxon potential fitted to the TCSM shapes.
The shell correction $\delta E(q)$  contains contributions from the shell effects within the independent particle (shell) model $\delta E_{shell}$  and in the pairing energy $\delta E_{pair}$ as shown in the equation
\begin{equation}
\label{deltae}
\delta E(q) =\sum_{n,p} \left( \delta E_{shell}^{(n,p)}(q) + \delta E_{pair}^{(n,p)} (q) \right) .
\end{equation}

In order to project the five-dimensional deformation energy onto the 2-dimensional surface we show in Fig. \ref{edef180} the mean value $\bra F\ket(z_0/R_0, \alpha)$ given by
\begin{equation}\label{Emean}
\bra F \ket(z_0/R_0, \alpha)=\frac{\sum_{\delta_1,\delta_2,\eps} F(q_{\mu})e^{-F(q_{\mu})/T_{coll}}}
{\sum_{\delta_1,\delta_2,\eps} e^{-F(q_{\mu})/T_{coll}}}
\end{equation}
Here $q_{\mu}$ is the set of parameters $q_{\mu}\equiv \{z_0/R_0, \delta_1,\delta_2, \alpha, \eps \}$ and for $T_{coll}$ we used the value $T_{coll}$= 2 MeV.

As one can see, the mean value of the potential energy $\bra F \ket(z_0/R_0, \alpha)$ is mass-symmetric both before and after the saddle. At the saddle there is a small bump with a symmetric shape that force trajectories to deviate from mass symmetry and leads to a mass-asymmetric yield.
\begin{figure}[ht]
\centering
\includegraphics[width=0.9\columnwidth]{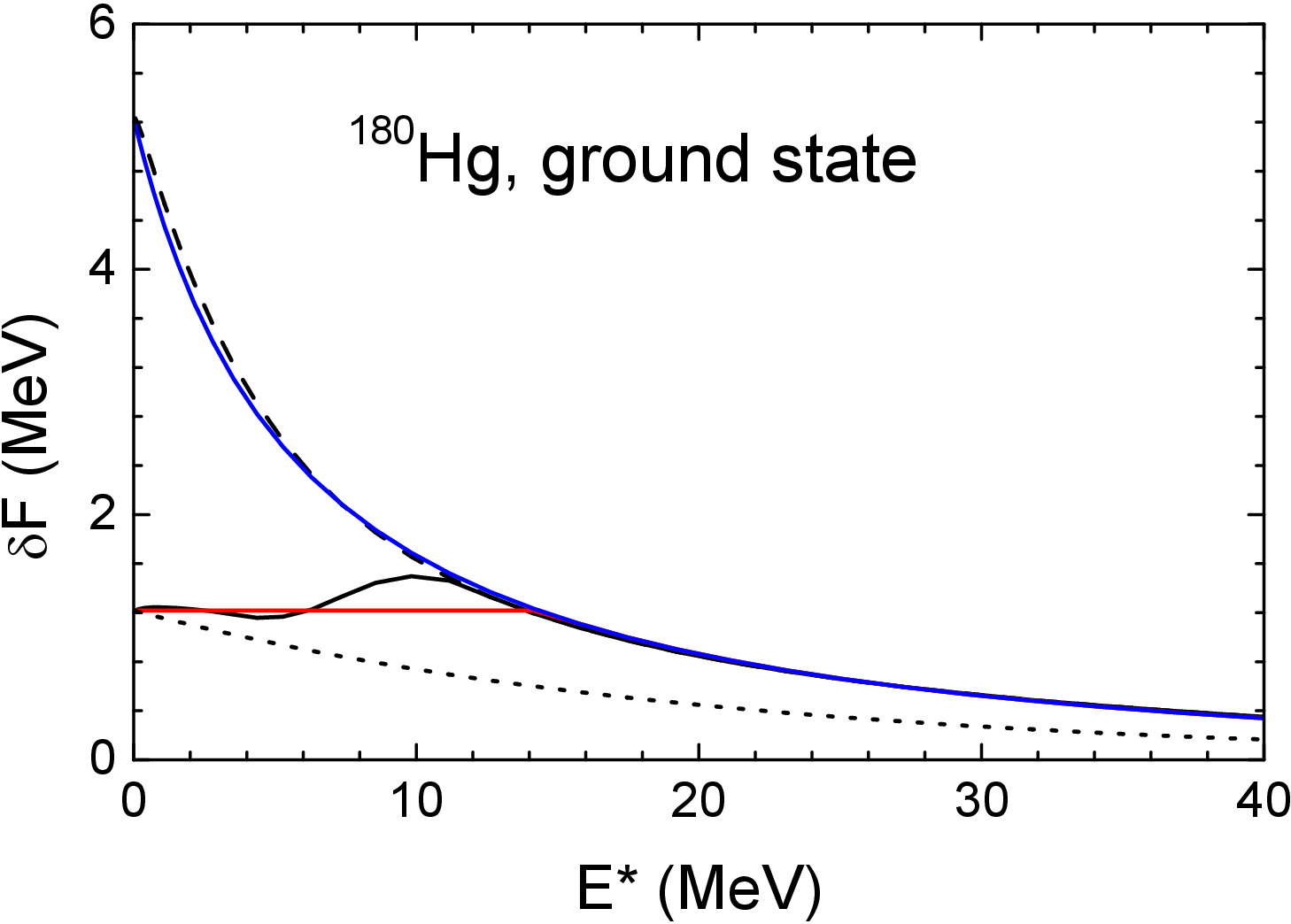}
\caption{(color online) The total shell correction $\delta F=\delta F_{shell}+\delta F_{pair}$ (\ref{deltafp}) (black solid line) and  $\delta F_{shell}$ (\ref{deltaft}) (dash) as a function of the excitation energy $E^*$ for $^{180}$Hs at the ground state. Red line - the approximation (\ref{delffit}), blue - the function (\ref{Phi}), and the dotted line is the exponential approximation (\ref{expon}) to $\delta F$. }
\label{shcota}
\end{figure}

For the energy $E(T)$ of system of independent particles at finite temperature one has
\begin{equation}\label{est}
E(T)=2\sum_k\epsk n_k^T, \, \text{with}\, n_k^T=\frac{1}{1+e^{(\epsk-\mu)/T}}.
\end{equation}
The averaged energy $\widetilde E(T)$ is defined by replacing the sum in (\ref{est}) by the integral with the smoothed density of states $\widetilde g(e)$
\begin{equation}\label{etavr}
\widetilde E(T)=\int_{-\infty}^{\infty}\widetilde g(e)e n_e^T de,
\end{equation}
with $n_e^T\equiv 1/[1+e^{(e-\tilde\mu)/T}]$.
The chemical potentials $\mu$ and $\tilde\mu$ in (\ref{est})-(\ref{etavr}) are defined by the particle conservation condition as
\begin{equation}\label{mues}
2\sum_k n_k^T=\int_{-\infty}^{\infty}\widetilde g(e) n_e^T de = N.
\end{equation}
The integrals in (\ref{etavr})-(\ref{mues}) should be calculated numerically.
The shell correction to the energy at finite temperature is then
\begin{equation}\label{deltaet}
\delta E_{shell}(T)=E(T)-\widetilde E(T).
\end{equation}
Another quantity of interest is the shell correction to the free energy given by
\begin{equation}\label{deltaft}
\delta F_{shell}(T)=\delta E_{shell}(T) -T \delta S_{shell}(T),
\end{equation}
For the entropy, the standard definition of $S(T)$ for the system of independent particles is used as
\begin{equation}\label{st}
S(T)=-2\sum_k [n_k^T \log n_k^T +(1-n_k^T)\log(1-n_k^T)].
\end{equation}
The average part of $S(T)$ is defined in an analogous way by replacing the sum in (\ref{st}) by the integral
\begin{equation}\label{stavr}
\widetilde S(T)=-\int_{-\infty}^{\infty}\widetilde g(e) [n_e^T \log n_e^T +(1-n_e^T)\log(1-n_e^T)] de .
\end{equation}
And the shell correction to the entropy is the difference between (\ref{st}) and (\ref{stavr}),
\begin{equation}\label{deltast}
\delta S_{shell}(T)=S(T)-\widetilde{S}(T).
\end{equation}
The calculated value of $\delta F_{shell}(T)$ at the ground state of $^{180}$Hg is shown by the dashed line in Fig. \ref{shcota}.

Next, we account for the pairing interaction in the Bar\-deen-Cooper-Schriffer (BCS) approximation \cite{bardeen}.
For the energy of independent quasi-particles at finite temperature one has,
\begin{eqnarray}\label{ebcst}
E_{BCS}(T)=2\sum_{k=k_1}^{k_2}\epsk n_k^{\Delta,T}-\frac{\Delta^2}{G},\quad\text{with}\nonumber\\
n_k^{\Delta,T}\equiv \frac{1}{2}\left(1-\frac{\epsk-\lambda}{E_k}\tanh\frac{E_k}{2T}\right),
\end{eqnarray}
where $G$ is the strength of the pairing interaction, $k_1$ and $k_2$ are the limits of the so called pairing window, $\Delta$ is the pairing gap and $E_k$ are the quasi-particle energies
\begin{equation}\label{eqp}
E_k=\sqrt{(\epsk-\lambda)^2+\Delta^2}.
\end{equation}

For the entropy one has the expression analogous to (\ref{st}),
\begin{equation}\label{sbcst}
S(T)=-2\sum_{k=k_1}^{k_2}[n_k^{\Delta,T} \log n_k^{\Delta,T} +(1-n_k^{\Delta,T})\log(1-n_k^{\Delta,T})]
\end{equation}
The pairing energy is then defined by the difference between (\ref{ebcst}) and the energy of independent particles within the pairing gap:
\begin{equation}\label{epairt}
E_{pair}(T)=E_{BCS}(T)-2\sum_{k=k_1}^{k_2} \epsk n_k^T\,.
\end{equation}
Similar, the pairing contribution to the entropy is
\begin{eqnarray}\label{spairt}
S_{pair}(T)=2\sum_{k=k_1}^{k_2} [n_k^T \log n_k^T +(1-n_k^T)\log(1-n_k^T)]\nonumber\\
-2\sum_{k=k_1}^{k_2}[n_k^{\Delta,T} \log n_k^{\Delta,T} +(1-n_k^{\Delta,T})\log(1-n_k^{\Delta,T})].
\end{eqnarray}
The average counterparts of $E_{pair}$ and $S_{pair}$ are defined by replacing the sum over quantal states $\vert k\rangle$ by the integrals with the average density of single-particle states  defined in terms of Strutinsky smoothing, see \cite{shcot}.
Then, the shell correction to the pairing energy is given by
\begin{equation}\label{deltafp}
\delta F_{pair}(T)=E_{pair}(T)-\widetilde E_{pair}(T)-T[S_{pair}(T)-\widetilde S_{pair}(T)].
\end{equation}
The  sum of $\delta F_{shell}+\delta F_{pair}$ at the ground state of $^{180}$Hg is shown by the solid line in Fig. \ref{shcota} as a function of the excitation energy $E^*=a T^2$.

The calculation of $\delta F$ at finite temperature is very time-consuming. To avoid lengthy computations one usually uses some parametrization of the dependence $\delta F$ on T or the excitation energy $E^*$. Most often the exponential approximation
\begin{equation}\label{expon}
\delta F(E^*)=\delta F(0) \exp(-E^*/\gamma),\, \text{with}\, \gamma\approx 20 \mev,
\end{equation}
is used, written in analogy to Ignatyuk's approximation
\cite{ignat} for the level density parameter $a$. This approximation works well for systems without pairing. In the presence of pairing correlations the dependence of $\delta F$ on $E^*$ is more complicated. At small excitations the variations of $\delta F_{shell}$ and $\delta F_{pair}$ almost cancel each other and the sum $\delta F=\delta  F_{shell}+\delta F_{pair}$ does not vary much with the excitation energy, as shown by the black solid curve in Fig. \ref{shcota}.

A more accurate approximation of the dependence $\delta F$ on $E^*$ was developed in \cite{shcot},
\begin{eqnarray}\label{delffit}
\delta F(E^*)=
 \delta F(0),\quad\text{if}\quad\vert\delta F_{shell}(0)\Phi(E^*)\vert\ge \vert\delta F(0)\vert\,, \nonumber\\
 \text{or}\quad\delta F_{shell}(0)\Phi(E^*),\,\text{if}\,\vert\delta F_{shell}(0)\Phi(E^*)\vert\leq \vert\delta F(0)\vert\,.\,\,
\end{eqnarray}
The function $\Phi(E^*)$ is given by
\begin{equation}\label{Phi}
\Phi(E^*)=(e^{-E_1/E_0-1})/(e^{(E^*-E_1)/E_0}-1),
\end{equation}
The parameters $E_0$ and $E_1$ were defined by fitting the calculated $\delta F_{shell}(E^*)$ by the function (\ref{Phi}).
The approximation for the values of $E_0$ and $E_1$ averaged over the mass number $A$ can be found in \cite{shcot}.

The most important conclusion from Fig. \ref{shcota} is that in the presence of pairing correlations  the shell correction $\delta F_{shell}+\delta F_{pair}$ decreases with the excitation energy much slower than the popular exponential approximation (\ref{expon}).
This may be the reason, why the influence of shell effects is observed in nuclear fission at relatively high excitation energies $E^* = 40\div 50$ MeV.
\section{The tensors of friction and inertia}
\label{transp}
In the present work we utilize the macroscopic transport coefficients commonly employed for solving the Langevin equations. These depend solely on the shape of the system and are independent of the excitation energy of the system.

The macroscopic mass tensor $M_{\mu\nu}^{WW}$ is typically defined in the Werner-Wheeler approximation \cite{werwhe} as:
\begin{equation}\label{massww}
M_{\mu\nu}^{WW} =
 \pi \rho_0 \int \rho^2(z, q) \left[ A_{\mu} A_{\nu}
+ \frac{\rho^2(z, q)}{8} A_{\mu}' A_{\nu}' \right] dz,
\end{equation}
where $\rho(z,q)$ is the profile function and
\begin{equation}
A_{\mu}(z;q)=\frac{1}{\rho^2(z,q)}\frac{\partial}{\partial q_{\mu}} \int_{z}^{z_R} \rho^2 (z',q) dz'.
\end{equation}
The shape of an axially symmetric nucleus is obtained by the rotation of the profile function $\rho(z,q)$ around z-axis.

The macroscopic friction tensor is determined by the so-called wall-and-window formula \cite{runswiat}.
The wall-and-window friction is a generalization of wall friction \cite{wall1} which for
axially symmetric shapes can be expressed as \cite{krapom},
\begin{equation}\label{frwall}
\gamma_{\mu\nu}^{\mathrm{wall}}=\pi\rho_{_0} \bar v\int dz \frac{\left(\frac{\partial \rho^2}{\partial q_{\mu}}+\frac{\partial \rho^2}{\partial z}\frac{\partial z_{cm}}{\partial q_{\mu}}\right)\left(\frac{\partial \rho^2}{\partial q_{\nu}}+\frac{\partial \rho^2}{\partial z}\frac{\partial z_{cm}}{\partial q_{\nu}}\right)}{\sqrt{4\rho^2(z,q)+(\partial \rho^2(z,q) / \partial z)^2}}\,,
\end{equation}
where $\rho_0=3mA/(4\pi R_0^3)$ and $v_F$ represents the Fermi velocity. The derivatives $\partial \rho^2(z,q)/\partial q$ in Eq.\,(5.172) of \cite{krapom} should be calculated under the condition that the nuclear center of mass remains fixed during the fission process.
 This condition is satisfied by including terms proportional to $\partial z_{cm}/\partial q$ in the numerator of (\ref{frwall}).

In the wall-and-window model the velocities of nucleons are considered relative to the velocities of the centers of mass of the left or right parts of the nucleus and a "window"  term is added \cite{runswiat,krapom},
\begin{equation}\label{frwall2}
\gamma_{\mu\nu}^{\mathrm{w+w}}=\pi\rho_0\bar v\left(\int_{z_{L}}^{0}I_L(z)\,dz+\int_{0}^{z_{_R}}I_R(z)dz \right)
+\gamma_{\mu\nu}^{\mathrm{window}},
\end{equation}
where
\begin{equation}\label{ILR}
I_{L,R}(z)=\frac{\left(\frac{\partial \rho^2}{\partial q_{\mu}}+\frac{\partial \rho^2}{\partial z}\frac{\partial z_{cm}(L,R)}{\partial q_{\mu}}\right)\left(\frac{\partial \rho^2}{\partial q_{\nu}}+\frac{\partial \rho^2}{\partial z}\frac{\partial z_{cm}(L,R)}{\partial q_{\nu}}\right)
}{\sqrt{4\rho^2(z,q)+(\partial \rho^2(z,q) / \partial z)^2}}
\end{equation}
and
\begin{equation}\label{frwindow}
\gamma_{\mu\nu}^{\mathrm{window}}=
\frac{1}{2}\rho_0\bar v\left[\Delta\sigma\left(\frac{\partial R_{12}}{\partial q_{\mu}}\frac{\partial R_{12}}{\partial q_{\nu}}\right)+\frac{32}{9\Delta\sigma}\frac{\partial V_L}{\partial q_{\mu}}\frac{\partial V_L}{\partial q_{\nu}}\right] \ ,
\end{equation}
where $R_{12}$ is the distance between the centers of mass of the left and right parts of the nucleus and $\Delta\sigma$ is the area of the "window".

One expects a smooth transition between the regime where the wall formula applies and the part of the fission path where wall-and-window friction should be used. For this Nix and Sierk \cite{nixsierk} proposed the phenomenological ansatz
\begin{equation}\label{total}
\gamma_{\mu\nu}^{total}=\sin^2(\pi\phi/2)\gamma_{\mu\nu}^{wall}+\cos^2(\pi\phi/2)\gamma_{\mu\nu}^{w+w},
\end{equation}
with $\phi=(r_{neck}/R_{min})^2$, where $R_{min}$ is the minimal semi-axis of the two outer ellipsoidal shapes.

Shortly after the introduction of wall friction it was noted that the wall friction is too strong and the reduction factor, $k_s=0.27$, was introduced by Nix and Sierk \cite{nix1987} based on the analysis of widths of giant resonances. This reduction factor was used in all our Langevin calculations.

\section{The fission fragments mass and total kinetic energy  distributions}
\label{results}
In this section we present the results of the numerical solution of the five-dimensional Langevin equations (1,2) with the specified potential energy and transport coefficients. In these calculations the damping of shell correction with the excitation energy was described using the method developed in \cite{shcot}. The frequency of local collective vibration was assumed to be constant, $\hbar\varpi$=2 MeV. In principle, $\varpi$ should depend on the deformation, but accounting for this dependence makes the computation time too long.
\begin{figure}[ht]
\centering
\includegraphics[width=0.95\columnwidth]{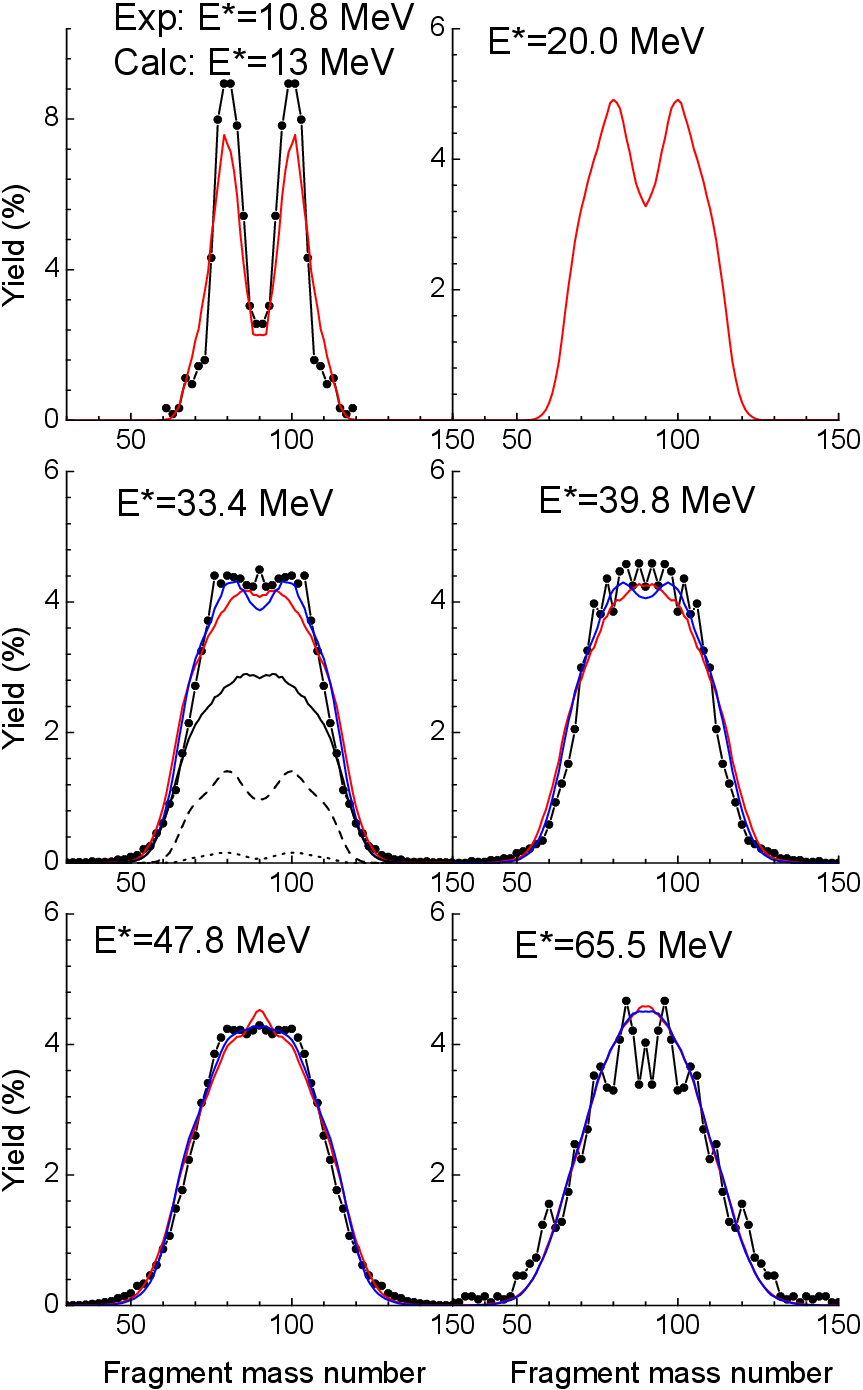}
\caption{(color online) The comparison of calculated fission fragment mass distributions (Eq.\ref{ffmd}, red lines) for $^{180}$Hg with the experimental data \cite{Nishio2015} (black lines with black points -– error bars are
omitted for legibility reason). Blue lines show the effect of multichance fission (\ref{yieldd}). For $E^*$=33.4 MeV, the contributions of first, second and third chance fission are shown by black solid, dashed and dotted lines, respectively.}
\label{yields180}
\end{figure}

For solving Eqs.(1,2), in addition to the coefficients of the equations, one should fix the initial values and final conditions. The calculations for all mercury isotopes were started from the same point close to the ground state deformation, $\{q_{\mu}\}=\{0.8, 0.0, 0.0, 0.0, 0.3\}$ with zero collective momenta $\{p_{\mu}\}=0$. The integration of the equations continued until the neck radius reached $r_{neck}^{(crit)}$=2 fm. At this point the solutions of the equations provide all the information on the shape, collective velocities and excitation energy of the system. This information enables the calculation of the moments of the density distribution, the mass distribution of fission fragments, the prescission and total kinetic energies, and the excitation energy of the nucleus just before scission.

The mass distribution of fission fragments $Y(A_F)$ is defined as the normalized to 200\% sum of contributions of all trajectories whose fragment mass $A_{Fi}$ at the scission point does not deviate from $A_F$ more than by 1/2 amu,
\begin{equation}\label{ffmd}
 Y(A_F)=200 \sum_{|A_{F\mu}-A_F|\leq 1/2}1\slash\sum_{A_F}\sum_{|A_{F\mu}-A_F|\leq 1/2}1.
\end{equation}
The $A_{F\mu}$ in Eq. (\ref{ffmd}) is the fission fragment mass at the deformation point $q_\mu$.
It is defined as $A_{F\mu} = 0.5 A_{CN} (1+\alpha_\mu)$,
where $A_{CN}$ is the mass number of the compound nucleus. By definition, $A_F$ is the integer value closest to $A_{F\mu}$.

The calculated fission fragment mass distributions for\-med in the reactions $^{36}$Ar +$^{144}$Sm $\Rightarrow$ $^{180}$Hg and
$^{36}$Ar + $^{154}$Sm $\Rightarrow$ $^{190}$Hg at few excitation energies are shown in Figs. \ref{yields180} and \ref{yields190} by red lines.
The black lines represent the experimental results from \cite{Nishio2015}. The agreement between the theory and experiment appears to be very  good.
\begin{figure}[ht]
\centering
\includegraphics[width=0.95\columnwidth]{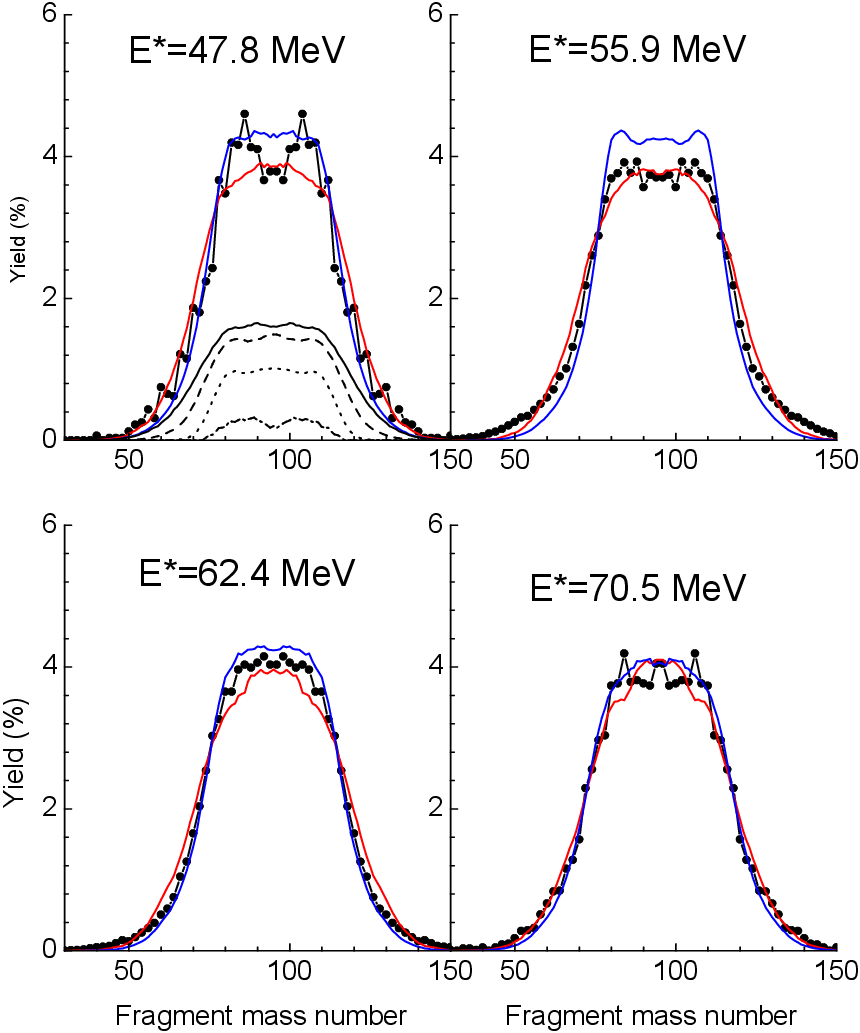}
\caption{(color online) The comparison of calculated fission fragment mass distributions (Eq.\ref{ffmd}, red lines) for $^{190}$Hg with the experimental data \cite{Nishio2015} (black lines with black points -– error bars are
omitted for legibility reason). Blue lines show the effect of neutron evaporation (\ref{yieldd}).
For $E^*$=47.8 MeV, the contributions of first, second, third and fourth chance fission are shown by black solid, dashed,  dotted and dashed-dotted lines, respectively.}
\label{yields190}
\end{figure}

We define the total kinetic energy $TKE$ as the sum of collective kinetic energy in the fission direction and the Coulomb repulsion energy between the point charges placed at the centers of mass of 'fragments' at the scission point
\begin{equation}
TKE=\frac{1}{2}(m^{-1})_{z_0z_0}p_{z_0}p_{z_0}+e^2\frac {Z_{1}Z_{2}}{D},
\end{equation}
where $D$ is the distance between the centers of mass of the left and right parts of nucleus at the scission point. The nucleus is divided into 'left' and 'right' by the position of the neck.

The distributions $P_{TKE}$ of fission events in total kinetic energy for the reactions $^{36}$Ar +$^{144}$Sm $\Rightarrow$ $^{180}$Hg and $^{36}$Ar +$^{154}$Sm $\Rightarrow$ $^{190}$Hg are presented in Fig. \ref{tkes}. The $P_{TKE}$ was defined as the number of fission events with the total kinetic energy within the interval $TKE \pm 1/2$ MeV.
\begin{figure}[ht]
\centering
\includegraphics[width=0.95\columnwidth]{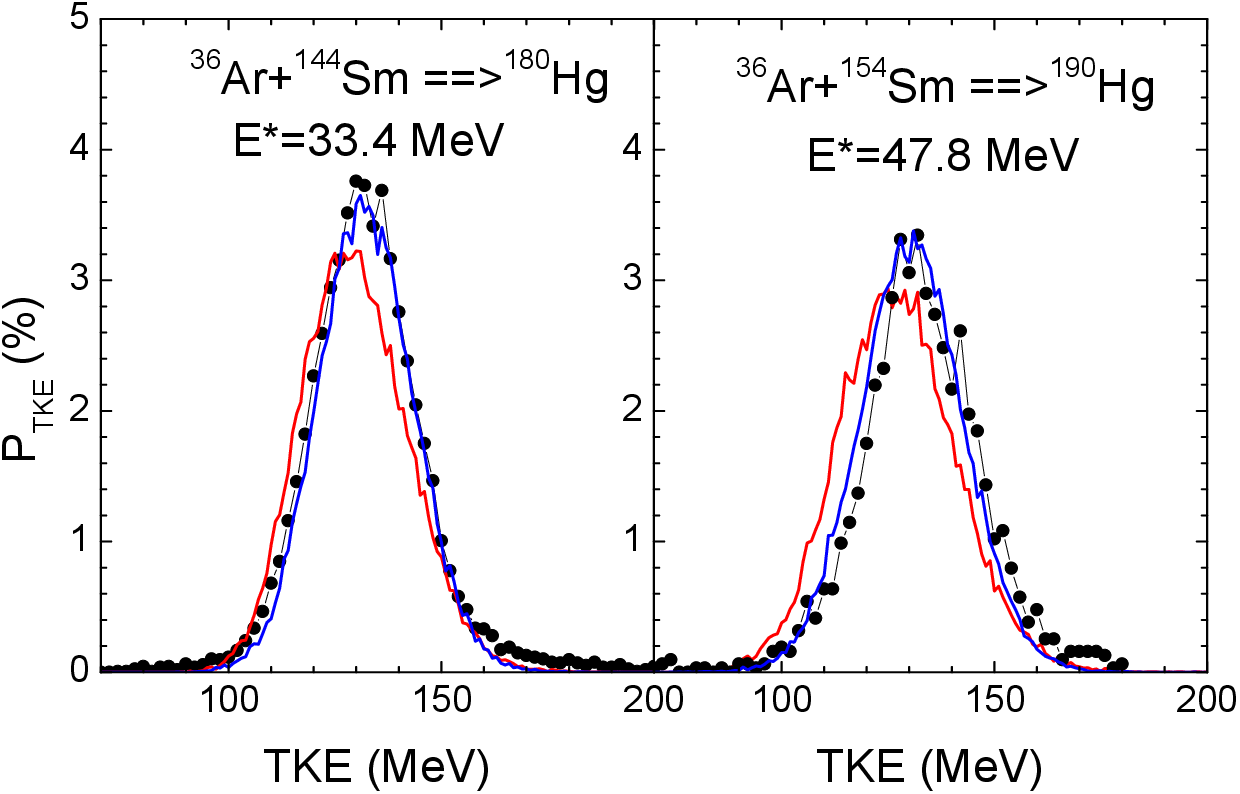}
\caption{(color online) The normalized to 100\% calculated distributions of fission events in total kinetic energy (red lines)  and the experimental data \cite{Nishio2015} (black lines - error bars are omitted for legibility reason). The red lines refer to the first-chance fission, and the blue lines refer to multi-chance fission (\ref{ptke}).}
\label{tkes}
\end{figure}

The agreement with experiment is reasonably good. Please note that all the calculations were carried out with the same set of "free" parameters, $k_s$=0.27, $r_{crit}$=2 fm, $\hbar\omega$= 2 MeV. No fitting was done.
\section{The multi-chance fission}
\label{multi}
The compound nuclei $^{180}$Hg or $^{190}$Hg
were produced at high excitation energies, between 30 and 70 MeV.  At such energies nuclei can emit particles before undergoing fission.
The fission after emission of charged particles is only a few percent of the total fission events.
The most probable emission is that of neutrons. Therefore, we consider  only neutron emission in the present work.

The fission probabilities $P_f$ after neutron emissions were calculated within the statistical Hauser-Feshbach  model \cite{hauser} by the TALYS code \cite{talys}. For the reduction of excitation energy due to neutron emission we used the approximation
\begin{equation}\label{Exx}
E^*(^{A-1}\text{Hg})=E^*(^{A}\text{Hg})-S_n(^{A}\text{Hg})-2T(^{A}\text{Hg}).
\end{equation}
Apart from the neutron separation energy $S_n$ the emitted neutron takes away a part of the nucleon's kinetic energy, approximately equal to $2T$, where
 $T=\sqrt{E^{*}/a}$, and $a$ is the level density parameter (\ref{alevel}).
For the neutron separation  energies $S_n$ and fission barriers we used the values calculated in \cite{Moller1997,Moller2015}.

Then, the total mass distribution $Y$ consists of contributions from multi-chance fission,
\begin{equation}\label{yieldd}
Y=\sum_{n=0,1,...}P_f(^{A-n}\text{Hg})Y(^{A-n}\text{Hg})\slash\sum_{n=0,1,...}P_f(^{A-n}\text{Hg}),
\end{equation}
with A=180 or A=190.

For the Hg isotopes involved in the different fission chances, we consider the same mass and inertia tensors, as they do not vary much with mass within a few units in the framework of the macroscopic prescriptions. On the contrary, the potential energy is recalculated for each chance, viz. it takes properly into account the change in neutron number of the fissioning isotope.

The number of terms in (\ref{yieldd}) depends on the initial excitation energy $E^*$. For larger $E^*$ the number of terms in (\ref{yieldd}) becomes larger. In the present calculations we accounted up to three subsequent neutron emission.

The mass distributions (\ref{yieldd}) are shown by the blue lines in Figs. \ref{yields180},\ref{yields190}. As one can see, the results of multi-chance fission do not differ much from the first chance fission.
The effect of multi-chance fission is seen only at small excitation energies.
This can be understood in the following way. The shape of all mass distributions (experimental and first-chance) is almost the same for all excitation energies. So, the admixture of the yields with smaller $E^*$ does not have a large effect on the yields. Only around $E^*$=20 MeV, and below, the shell effects are clearly seen. However, the fission width at small excitations is very small, and the contribution of the third-, fourth-chance fission to the total yield is also small.
Besides, due to the competition of fission and neutron evaporation, the probability of fission gets additionally smaller with smaller excitation energy.

In case of $^{180}$Hg at $E^*$=33.4 MeV the calculated mass distribution shows somewhat more shell effects compared with experimental results.

We have estimated also the contributions of multi-chance fission to the distribution of fission events in kinetic energy
\begin{eqnarray}\label{ptke}
P_{TKE}&=&\sum_{n=0,1,2,...}P_f(^{A-n}\text{Hg})P_{TKE}(^{A-n}\text{Hg})\nonumber\\
&\slash&\sum_{n=0,1,2,...}P_f(^{A-n}\text{Hg}),
\end{eqnarray}
where $P_{TKE}(^{A-n}$Hg) are the TKE distributions calculated for the reduced excitation energies.

The values $P_{TKE}$ (\ref{ptke}) are shown by blue curves in Fig. \ref{tkes}.
One can see that the account of multi-chance fission has
a noticeable effect on the TKE distributions, and leads to
an extremely good description of the measurements. While the
influence of post-scission evaporation (by the fragments after scission) is well
known and has been widely discussed, the influence of pre-scission evaporation
(by the compound nucleus before scission) has been rarely addressed. The present
work demonstrates its non-negligible influence.
\begin{figure}[ht]
\centering
\includegraphics[width=0.9\columnwidth]{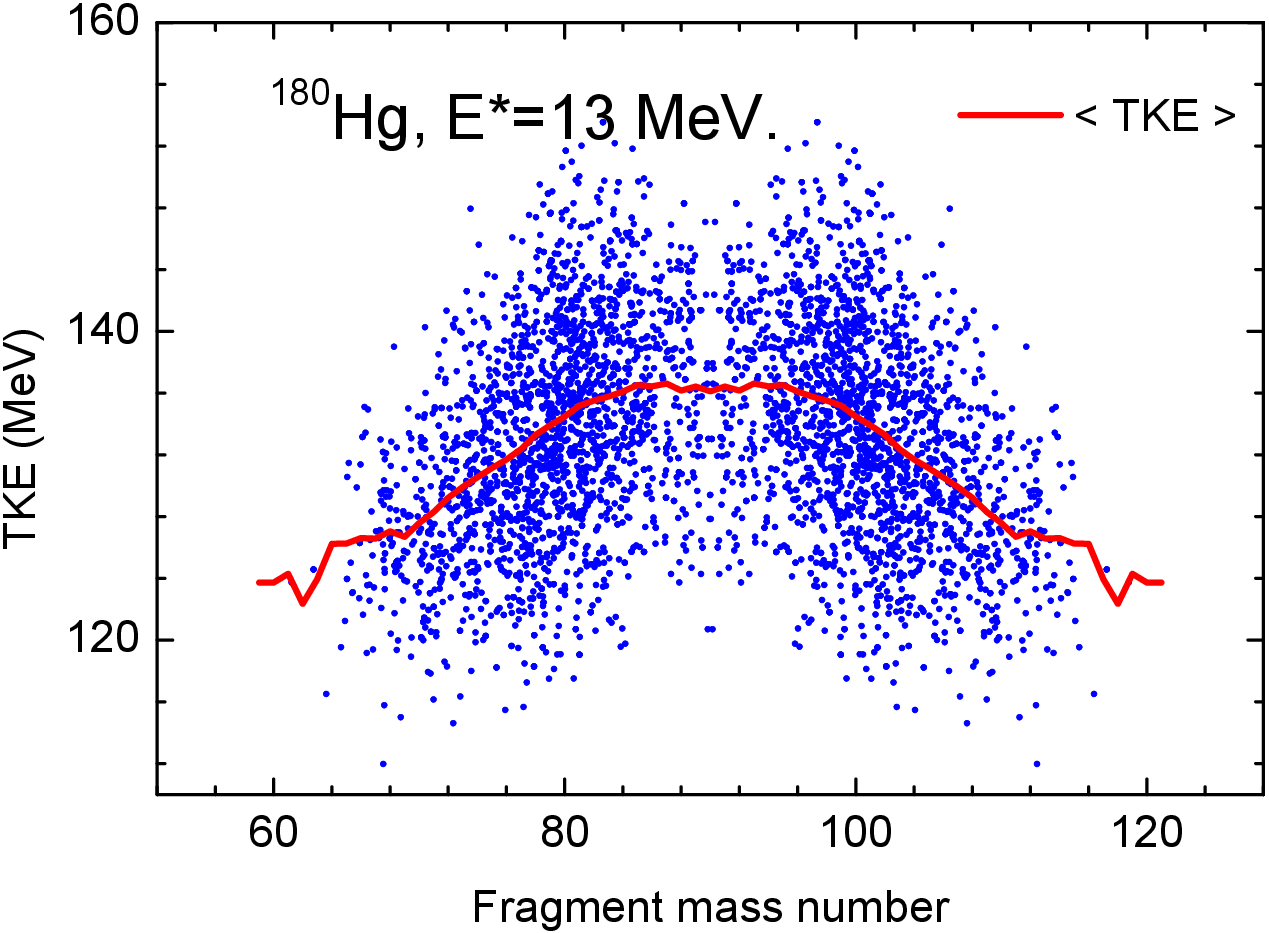}
\caption{(color online) The distribution of fission events in TKE - mass asymmetry plane. The average value of TKE is shown by red solid line.}
\label{tke13}
\end{figure}
Figure \ref{tke13} presents the correlation between mass number and total kinetic energy (TKE) of the fission fragments of $^{180}$Hg at an excitation energy of 13 MeV. In the case of the uranium region, it is known \cite{andrey} that the asymmetric
fission region has higher TKE due to the so-called Standard modes, S1 and S2, driven by the shell effects \cite{andrey,scamps}, in addition to a Superlong symmetric mode (liquid drop model component) characterized by lower TKE. However,
for $^{180}$Hg, it is observed that only an asymmetric fission component exists. This suggests that the asymmetric and symmetric fission pathways originating from the compound nucleus are separated by a higher ridge in $^{180}$Hg compared to the
uranium region. The complete absence of the Superlong mode is unexpected and constitutes an important finding
for understanding the fission mechanism.

Figure \ref{Q20a} shows the average quadrupole moment $\langle Q_{20} \rangle$ as a function of fission fragment mass number for a few values of the excitation energy. This quantity provides us with general information about the shape of the fission fragments. From the figure, it can be seen that this quantity tends to increase overall as the fragment mass number increases. 
In other words, it shows that in the fission of $^{180}$Hg the heavy fragments are produced in a more deformed shape than their light counterparts. 
Interestingly, it can be seen that there is a kink in the symmetric fission component at low excitation energies of 13 and 20 MeV. Here, fission fragments near $A_F$=90, namely $^{90}$Zr, are produced in a more compact shape compared to fragments having nearby mass numbers. But, such peculiarity seems to wash out as the excitation energy increases. 
Furthermore, although the shell effect at $A_F$=90 has a strong influence on the shape of the fission fragments, it does not seem to give any enhancement in terms of the mass number distribution.  Such independence of the "shape" and "mass distribution" of fission fragments seems to be a unique characteristic of the scission mechanisms of  $^{180}$Hg isotope.


\begin{figure}[ht]
\centering
\includegraphics[width=0.88\columnwidth]{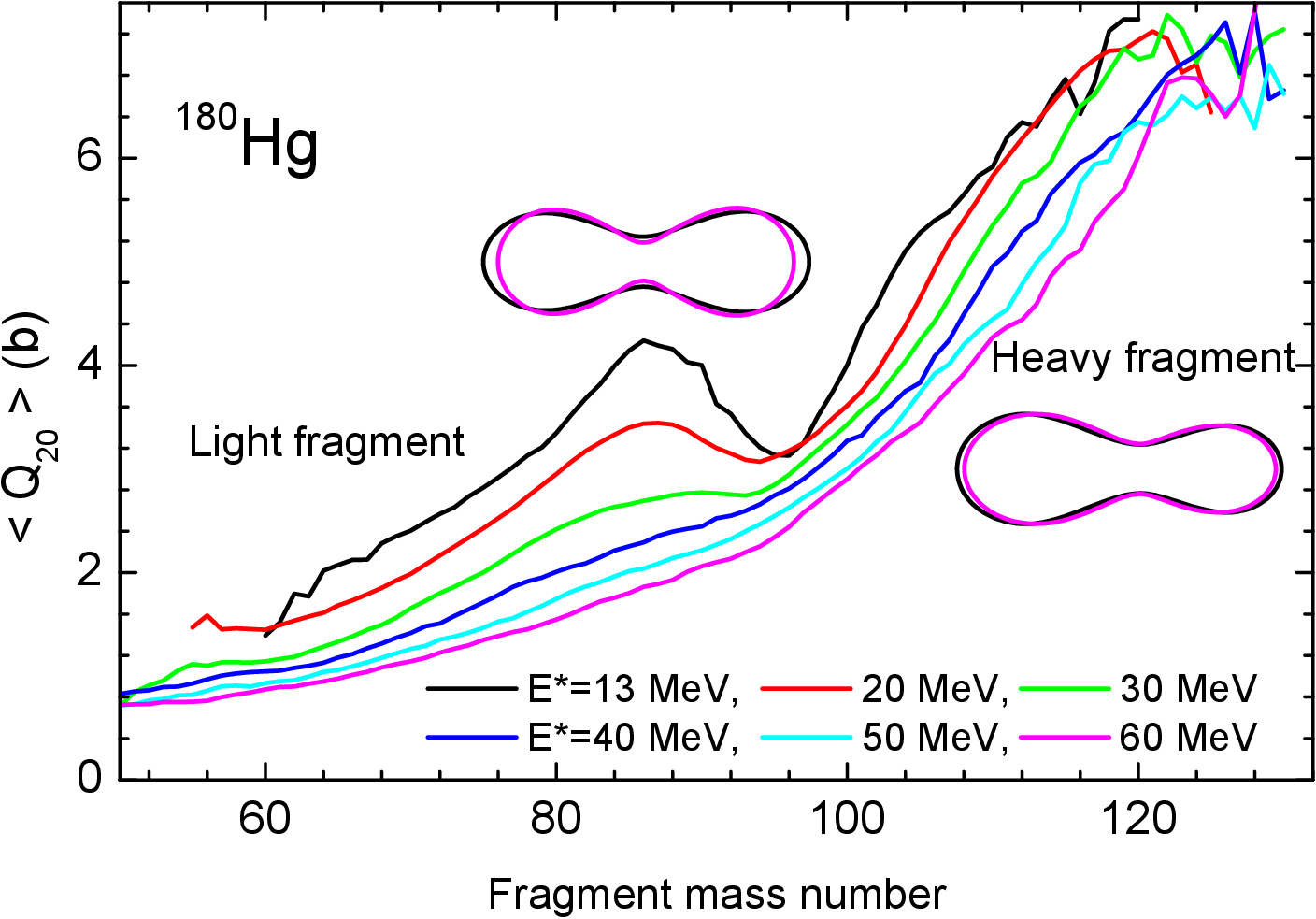}
\caption{(color online) The quadrupole moments of light and heavy parts of $^{180}$Hg just before scission. The values corresponding to different excitation energies are shown by different colors. Two examples of representative scission shapes for  ($A_L, A_H$)=(86, 94) and ($A_L, A_H$)=(60, 120) at $E^*$=13 MeV and $E^*$=60 MeV are shown by black and pink contours, respectively.}
\label{Q20a}
\end{figure}
It shall be noted that the quadrupole deformation $Q_{20}$ does not suffice to fully characterize a shape. This is illustrated in Fig. \ref{Q20a} 
where two examples of scission shapes having the largest variation
of quadrupole moment with the excitation energy are shown, namely for  ($A_L, A_H$)=(86, 94) and ($A_L, A_H$)=(60, 120) for $E^*$=13 MeV and $E^*$=60 MeV. The variation of the scission shapes with the $E^*$ at fixed mass asymmetry is rather small. Thus, at larger excitation energy the fissioning shape of $^{180}$Hg becomes slightly shorter.
\section{Summary}
\label{summa}
The fission fragments mass and kinetic energy distributions have been calculated for the fission of  $^{180}$Hg and $^{190}$Hg formed  in the reactions $^{36}$Ar +$^{144}$Sm $\Rightarrow$ $^{180}$Hg and
$^{36}$Ar +$^{154}$Sm $\Rightarrow$ $^{190}$Hg at few values of the initial excitation energy. The calculations were carried out within the recently developed five-dimensional Langevin approach for the description of fission of heavy nuclei. The main development of the approach was a more accurate treatment of the collisions of fission trajectories with the boundaries of the deformation space. The so called soft wall was placed at the boundaries to keep the trajectories within the deformation space.

Special attention was paid to the accurate description of the dependence of shell effects on the excitation energy. Instead of widely used Ignatyuk's prescription, we used more accurate approximation developed in \cite{shcot}.

Preliminary calculations were carried out for the so called first chance fission, for which neutron
emission by the compound nucleus prior scission does not occur. The modeling was then enhanced to
account for multichance fission with the  probability increasing with the bombarding energy. Emission of up
to three neutrons before scission was considered.

The fission fragment mass distributions are reproduced surprisingly well for both $^{180}$Hg and $^{190}$Hg for all considered excitation energies. The effect of multi-chance fission on the mass distribution is very small. The results accounting for first chance only are already quite close to the experimental yields.

The effect of neutron emission on the kinetic energy
distribution has been found to be noticeable. Accounting for multichance
fission leads to very good agreement with the experimental data.

While all previous
work intended to point out the manifestation of multichance fission based on mass distributions,
our work strongly suggests that evaporation by the compound nucleus on the pre-scission TKE has
a measurable impact, and that the TKE is a very valuable observable to be investigated to
study multichance fission. For cases like this, it may be of precious value.

As explained in the introduction, we do not assume any
distribution as the initial condition for the five dynamical variables.  Our calculation starts from $\delta$-functions in all  five variables as the initial distributions.
Therefore, the positions and shapes of distributions in both mass and TKE arise as results of the time evolution described by our dynamical model.  This shows that the fluctuation and dissipation nature of the theory, which conforms the essential part
 of our approach, is vital characteristics to understand mechanisms of nuclear fission.

The calculations presented in this work confirm that our five-dimensional Langevin approach is a reliable tool for the theoretical predictions of the fission process observables.

{\bf Acknowledgments.}
The authors would like to express their gratitude to Profs. K. Nishio and K. Pomorski for valuable comments and suggestions.

\end{document}